\documentclass[11pt,twoside]{article}

\usepackage{asp2006}
\usepackage{epsf}
\usepackage{psfig}
\usepackage{lscape}
\usepackage{graphicx}
\usepackage{multicol}

\def\kms{\hbox{km s$^{-1}$}}
\def\VLSR{\hbox{$V_{\rm LSR}$}}

\def\sun{\hbox{$\odot$}}
\def\R32/10{\hbox{$R_{3\mbox{--}2/1\mbox{--}0}$}}

\begin{document}
\title{The ASTE Galactic Center CO {\it J}=3--2 Survey: \\ Probing Shocked Molecular Gas in the CMZ}   


\author{Tomoharu Oka\altaffilmark{1}, Kunihiko Tanaka\altaffilmark{1}, Shinji Matsumura\altaffilmark{1}, Makoto Nagai\altaffilmark{2}, Kazuhisa Kamegai\altaffilmark{3}, and Tetsuo Hasegawa\altaffilmark{4}}   
\altaffiltext{1}{\it Department of Physics, Faculty of Science and Technology, Keio University, 3-14-1 Hiyoshi, Kohoku-ku, Yokohama, Kanagawa 223-8522, Japan}    
\altaffiltext{2}{\it Institute of Physics, University of Tsukuba, 1-1-1 Ten-nodai, Tsukuba, Ibaraki 305-8571, Japan}    
\altaffiltext{3}{\it Institute of Space and Astronautical Science, Japan Aerospace Exploration Agency, 3-1-1 Yoshinodai, Sagamihara, Kanagawa, 229-8510 Japan}    
\altaffiltext{4}{\it Joint ALMA Observatory, El Golf 40, Piso 18, Las Condes, Santiago, Chile}    

\begin{abstract} 
Large-scale CO surveys have revealed that the central molecular zone (CMZ) of our Galaxy is characterized by a number of expanding shells/arcs, and by a peculiar population of compact clouds with large velocity widths --- high-velocity compact clouds (HVCCs).  We have performed a large-scale CO {\it J}=3--2 survey of the CMZ from 2005 to 2008 with the Atacama Submillimeter-wave Telescope Experiment (ASTE).  The data cover almost the full extent of the CMZ and the Clump 2 with a 34\arcsec\ grid spacing.  The CO {\it J}=3--2 data were compared with the CO {\it J}=1--0 data taken with the Nobeyama Radio Observatory 45 m telescope.  Molecular gas in the CMZ exhibits higher {\it J}=3--2/{\it J}=1--0 intensity ratios ($\R32/10\!\sim\!0.7$) than that in the Galactic disk ($\R32/10\!\sim\!0.4$).  We extracted highly excited, optically thin gas by the criterion, $\R32/10\!\geq\!1.5$.  Clumps of high \R32/10\ gas were found in the Sgr A and Sgr C regions, near SNR G 0.9+0.1, and three regions with energetic HVCCs; CO 1.27+0.01, CO --0.41--0.23, and CO --1.21--0.12.  We also found a number of small spots of high \R32/10\ gas over the CMZ.   Many of these high \R32/10\ clumps and spots have large velocity widths, and some apparently coincides with HVCCs, suggesting that they are spots of shocked molecular gas.  Their origin should be local explosive events, possibly supernova explosions.  These suggest that the energetic HVCCs are associated with massive compact clusters, which have been formed by microbursts of star formation.  Rough estimates of the energy flow from large to small scale suggest that the supernova shocks can make a significant contribution to turbulence activation and gas heating in the CMZ.   
\end{abstract}



\section{Introduction}
The central molecular zone (CMZ) of our Galaxy contains large amount of warm ($T_{\rm k}=30$--$60$ K; Morris et al. 1983) and dense [$n({\rm H}_{2})\geq 10^{4}$ cm$^{-3}$; Paglione et al. 1998] molecular gas.  The gas temperature is considerably higher than the dust temperature ($T_{\rm d}\!\simeq\!20$ K; Pierce-Price et al. 2000), suggesting that a gas heating process unrelated to ultraviolet radiation is working there.  Molecular clouds in the CMZ generally have large velocity widths ($\Delta V\!\geq\!15$ \kms).  They seem to be in equilibrium with external pressure, not bound by self-gravity (Oka et al 2001a).  Refractory molecules such as SiO (Mart\'in-Pintado et al. 1997; H\"uttemeister et al. 1998) and complex organic molecules (COMs; Requena-Torres et al. 2006) are common in the CMZ.  These suggest that interstellar shock pervades over the CMZ.   The origin of pervasive shock is still unclear.  Although theories of large-scale shocks caused by orbit crowding (Rodr\'iguez-Fern\'andez et al. 2006) or magnetic floatation (Fukui et al. 2008) can describe it, the other possibilities cannot be ruled out.  A `classical' interpretation, the accumulation of small-scale shocks by supernova explosions, might be able to contribute as well, however.  

About fifteen years ago, we have made a large-scale CO {\it J}=1--0 survey of the Galactic center region using the Nobeyama Radio Observatory (NRO) 45 m telescope (Oka et al. 1998).  This survey completely covers the CMZ and the Clump 2 (Bania 1977) with a 34\arcsec\ grid spacing.  High-resolution CO images have shown that the distribution and kinematics of molecular gas are highly complex, and that the morphology is characterized by filamentary structures and by a number of shells and/or arcs.  Some of the shells/arcs show clear expanding motion, and two of them, an expanding barrel adjacent to the nonthermal vertical filaments of the radio arc (Oka et al 2001b) and an expanding arc in the Galactic-north of Sgr B1 (Tanaka et al 2009), have already been reported and discussed in separate papers.

\begin{figure}[htbp]
\begin{center}
\includegraphics[width=11cm]{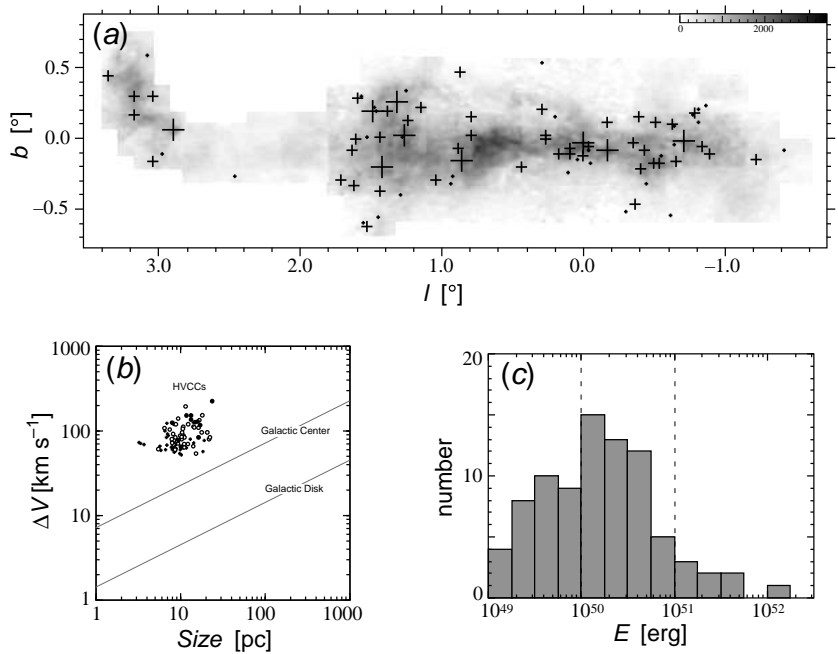}
\end{center}
\caption{({\it a}) Spatial distribution of HVCCs (Nagai 2008) superposed on the CO {\it J}=1--0 integrated intensity map.  Size of each cross shows the magnitude of kinetic energy.  ({\it b}) Size versus velocity-width plot for HVCCs, with the best-fit lines for normal Galactic center clouds and for clouds in the Galactic disk.  ({\it c}) Frequency distribution of the kinetic energies of HVCCs.  }\label{fig1}
\end{figure}

The most important result of the NRO 45m CO {\it J}=1--0 survey may be the discovery of high-velocity compact clouds (HVCCs).  At present, this population of molecular clouds is peculiar to the central region of our Galaxy.  HVCCs are defined by their compact sizes and large velocity widths, which lead to their short expansion times and enormous kinetic energies.  Two energetic HVCCs, CO 0.02--0.02 (Oka et al. 1999) and CO 1.27+0.01 (Oka et al. 2001c) have been studied in details.  We developed an identification scheme of HVCCs and identified 84 HVCCs in the CO {\it J}=1--0 data set (Nagai 2008).  They spread over the CMZ and Clump 2, but very few in the 200-pc molecular ring (Kaifu et al. 1972; Scoville 1972).  Some of them are isolated, but most of them are associated with giant molecular clouds.  The size versus velocity-width plot for HVCCs is well apart from the best-fit lines for normal Galactic center clouds and for clouds in the Galactic disk, indicating that they belong to a category different from normal clouds.  Their kinetic energies range from $10^{49}$ to $10^{52}$ erg, and about 2/3 of them exceed $10^{50}$ erg.  

These characteristics of HVCCs suggest that each of them have been formed by local explosive events, possibly a series of supernovae or hypernovae.  Therefore, energetic ones must be associated with massive stellar clusters.  In fact, an energetic HVCC CO 0.02--0.02 is associated with an emission cavity, which contains a group of mid-infrared point-like sources (Oka et al. 2008).  These HVCCs must be important targets not only because they have peculiar characteristics, but also because they could be related to the origin of pervasive shock, which is responsible for turbulence activation and gas heating in the CMZ.  In addition, they could be an indirect tracer of massive stellar clusters which are severely veiled in dense dusty material.

\section{ASTE Observations}
Since 2005, we have been performing a large-scale CO {\it J}=3--2 (345.795989 GHz) survey of the Galactic center region using the Atacama Submillimeter Telescope Experiment (ASTE).  ASTE is a Japanese 10 m submillimeter-wave telescope built at Pampa la Bola (alt.4800m), Chile (Kohno et al. 2004).  Objectives of this survey are, (1) to trace shocked molecular gas, (2) to unveil the origin of pervasive shock, and (2) to trace star formation history, in the Galactic center region.  Preliminary results of this survey have already been published (Oka et al. 2007), and the full presentation of the survey will be published in the forthcoming paper (Nagai et al. 2010).  

The telescope is equipped with a 345 GHz SIS receiver SC345.  The receiver had an IF frequency of 6.0 GHz, and the local oscillator was centered at 339 GHz.  The telescope has a beam size of 22\arcsec\ (FWHM) and a beam efficiency of 0.6 at 346 GHz.  The observations were carried out July 19--25 2005 , July 21--August 1 2006, and April 8--June 2 2008.  About 25,000 CO {\it J}=3--2 spectra have been collected with a 34\arcsec\ grid spacing, which is the same as the NRO 45m CO {\it J}=1--0 survey, covering the bulks of the CMZ and Clump 2.  All spectra were obtained with an XF-type autocorrelation spectrometer, which was operated in the 512 MHz bandwidth (1024 channel) mode.  The observations were performed by position-switching to a clean reference position, $(l, b)$=$(0\deg, -2\deg)$.  The on-source integration time was 10 seconds for each position and rms noise was 0.3 K in the $T_A^*$ scale.

\begin{figure}[htbp]
\begin{center}
\includegraphics[width=11cm]{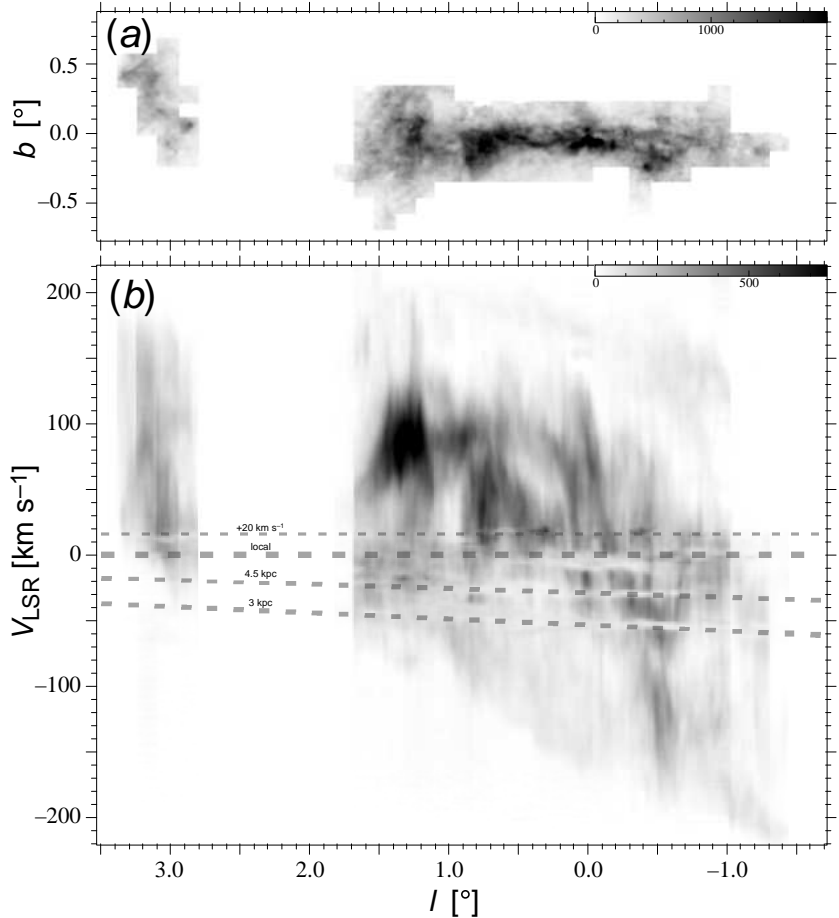}
\end{center}
\caption{({\it a}) Integrated intensity ($\int T_{\rm MB}\,dV$) map of CO {\it J}=3--2 emission.  The emission is integrated over velocities within $|V_{\rm LSR}|\!\leq\!200$ \kms.   ({\it b}) Longitude-velocity map of CO {\it J}=3--2 emission integrated over the observed latitudes ($\sum T_{\rm MB}$).  The data have been smoothed to a 60\arcsec\ resolution and summed up to each $+2$ km s$^{-1}$ bin.  ({\it c}) Frequency distributions of \R32/10\ weighted by the CO {\it J}=1--0 intensity.  }\label{fig2}
\end{figure}

\section{Results and Analyses}
\subsection{Intensities and Intensity Ratio}
As a natural result, spatial and longitude-velocity distributions of the {\it J}=3--2 emission are similar to those of the {\it J}=1--0 emission.  The main ridge of the CMZ is more intense in the {\it J}=3--2 emission, while the 200-pc molecular ring and Clump 2 are less prominent.  Four spiral arms in the Galactic disk, 3 kpc, 4.5 kpc, local, and +20 km s$^{-1}$ arms are seen as absorption features.  

We often refer ratios between molecular line intensities in diagnoses of physical conditions or chemical compositions of interstellar molecular gas.  The total CO {\it J}=3--2/CO {\it J}=1--0 intensity ratio was 0.73, which is higher than that of the disk gas ($\R32/10\!\sim\!0.4$; Oka et al. 2007), and is lower than that of the central region of M83 ($\R32/10\!\simeq\!1$; Muraoka et al. 2007).  The frequency distribution of \R32/10\ weighted by CO {\it J}=1--0 intensity shows a peak at 0.75.

\begin{figure}[htbp]
\begin{center}
\includegraphics[width=11cm]{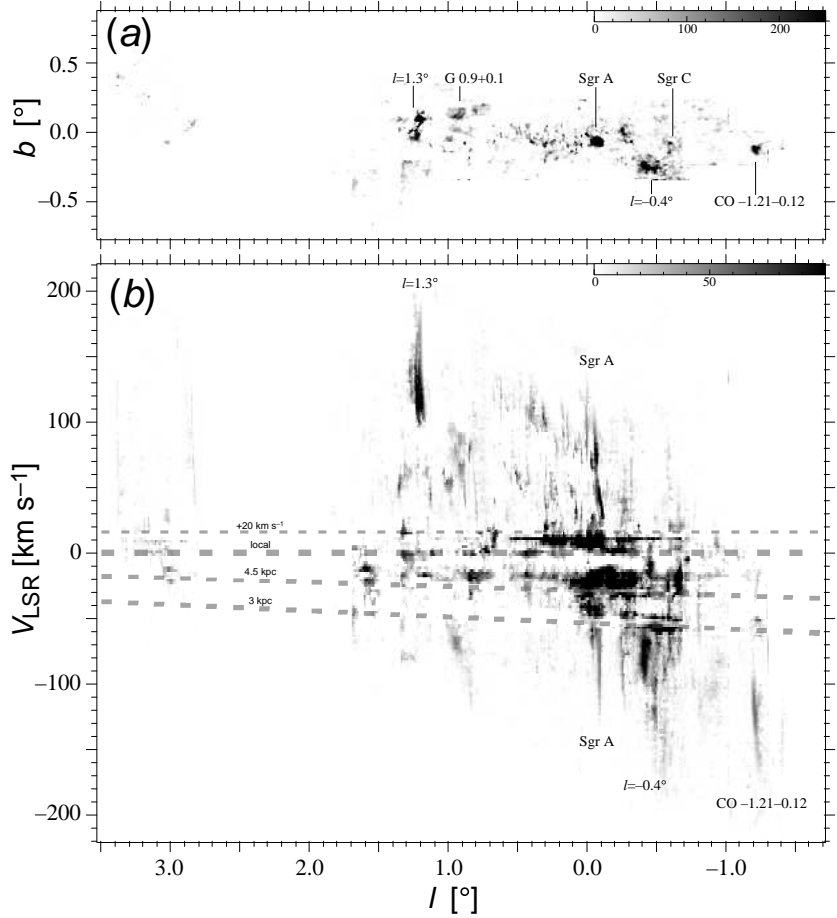}
\end{center}
\caption{({\it a}) Integrated intensity ($\int T_{\rm MB}\,dV$) map of CO {\it J}=3--2 emission for data with $\R32/10\!\geq\!1.5$.  The velocity range severely contaminated by the foreground disk gas [$(-55\!+\!10\,l)\,\kms\!\leq\!V_{\rm LSR}\!\leq\!+15\,\kms$] have been excluded.  ({\it b}) Longitude-velocity map of CO {\it J}=3--2 emission integrated over the observed latitudes ($\sum T_{\rm MB}$) for data with $\R32/10\!\geq\!1.5$.  ({\it c}) Frequency distributions of \R32/10\ weighted by the CO {\it J}=1--0 intensity.  ({\it d}) Curves of \R32/10\ as a function of $n({\rm H_2})$ and $T_{\rm k}$ for $N_{\rm CO}/dV\!=\!10^{17}$ cm$^{-2}$ (km s$^{-1}$)$^{-1}$.  }\label{fig3}
\end{figure}

\subsection{High $R_{3\mbox{--}2/1\mbox{--}0}$ Clumps}
Generally, \R32/10\ can be a measure of the kinetic temperature ($T_{\rm k}$) and the molecular hydrogen density [$n({\rm H}_2)$] if the CO column density per unit velocity width ($N_{\rm CO}/dV$) is not very high.  High \R32/10\ ratios have been found in UV-irradiated cloud surfaces near early-type stars (e.g., White et al. 1999), and shocked molecular gas adjacent to supernova remnants (e.g., Dubner et al. 2004).  \R32/10\ exceeds unity when the {\it J}=3 level is well thermalized and the lines are less opaque.  Here we try to extract highly excited gas from the CO data sets by $\R32/10\!\geq\!1.5$.   One-zone LVG calculations say that $\R32/10\!\geq\!1.5$ corresponds to $n({\rm H}_2)\!\geq\!10^{4}$ cm$^{-3}$ and $T_{\rm k}\!\geq\!50$ K when $N_{\rm CO}/dV\!=\!10^{17}$ cm$^{-2}$ (km s$^{-1}$)$^{-1}$.  These parameters are typical of shocked molecular gas.

We found several clumps of high \R32/10\ gas in the Sgr A and Sgr C regions, near SNR G 0.9+0.1, and three regions with energetic HVCCs; CO 1.27+0.01 ($l\!=\!1.3\deg$), CO --0.41--0.23 ($l\!=\!-0.4\deg$), and CO --1.21--0.12.  Note that most of these high \R32/10\ clumps have extremely large velocity widths.  Needless to say, these high \R32/10\ clumps are targets of great interest.  Two of them have been researched in more details:   

\subsubsection{Infalling Disk of High \R32/10\ Gas: }
In the Sgr A region, we notice an ellipse in the longitude-velocity map.   This seems to be a pretty large rotating disk of high \R32/10\ gas with a radius of about 8 pc (Oka et al. 2007, 2010).  It apparently has radial motion.  The upper segment of the disk is redshifted, while the lower segment is blueshifted.  This disk seems to include the well-known structure, the circumnuclear disk (CND; Genzel et al. 1985; G\"usten et al. 1987).  We know the upper part of the CND is in near side (e.g., Karlsson et al. 2003).  If we assume the continuity between the high \R32/10\ disk and the CND, the radial motion of the disk should be infalling.  The mass of the high \R32/10\ disk is $(2\mbox{--}6)\!\times\!10^5\,M_{\sun}$.  The rotation velocity of the disk is about 110 \kms\ and the infall velocity is 30--60 \kms.  Thus the infall timescale is comparable to the rotation period, indicating this disk must be transient.

\begin{figure}[ht]


\begin{center}
\includegraphics[width=11cm]{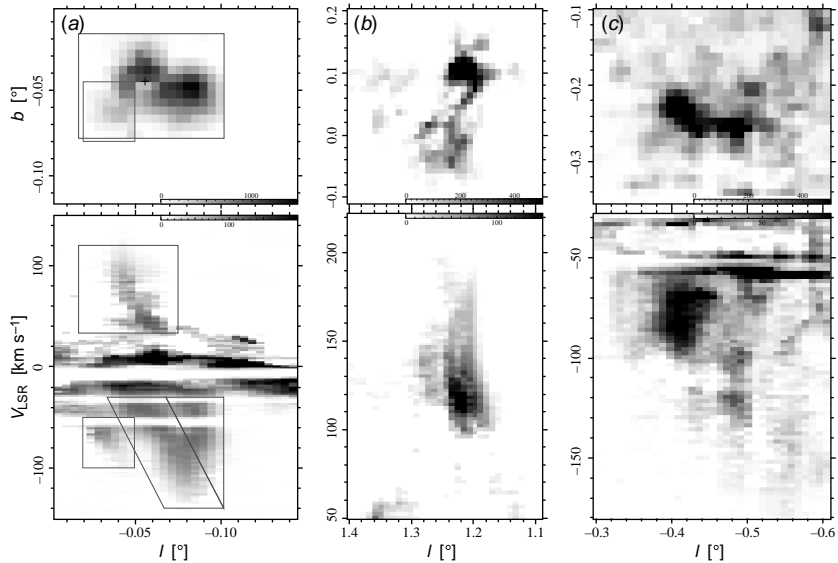}
\end{center}

\caption{({\it a}) Spatial and longitude-velocity distribution of high \R32/10\ gas in the Sgr A region.  Gray frames show the integration ranges.  The cross denotes the position of Sgr A$^*$.
({\it b}) Spatial and longitude-velocity distribution of high \R32/10\ gas in the $l\!=\!1.3\deg$ region.  The velocity integration range is $\VLSR\!=\!50$ to $220$ \kms.  
({\it c}) Spatial and longitude-velocity distribution of high \R32/10\ gas in the $l\!=\!-0.4\deg$ region.  The velocity integration range is $\VLSR\!=\!-180$ to $-30$ \kms.  
}\label{fig4}

\end{figure}

\subsubsection{Discovery of a Proto-superbubble: }
The $l\!=\!1.3\deg$ region corresponds to the center of a large molecular complex with a prominent elongation toward positive Galactic latitude (Oka et al 1998).  Two expanding shells, which have been identified in the CO {\it J}=1--0 data (CO 1.27+0.01; Oka et al. 2001c), appear as a high \R32/10\ clump.  We have made follow-up observations of this region using the NRO 45 meter telescope (Tanaka et al. 2007).  We detected intense, high-velocity  HCN and HCO$^{+}$ emission from this region, and found at least 9 expanding shells of dense molecular gas.  We also detected three isolated SiO clumps in the velocity ends of the expanding shells.  Each of the expanding shell has a kinetic energy of $10^{50.7\mbox{--}52.0}$ erg, and an expansion time of $10^{4.6\mbox{--}5.3}$ yrs.  The total kinetic energy amounts to $\sim\!10^{53}$ erg.  The association of dense/hot gas and SiO clumps and the enormous kinetic energy of the expanding shells ensure that they are generated by a series of supernova explosions, suggesting that a microburst of star formation has occurred there in the recent past.  The situation is similar to those of molecular superbubbles found in nearby starburst galaxies.  For instance, molecular superbubbles in NGC 253 have kinetic energies of $\sim\!10^{53}$ erg, and ages of $\sim\!0.5$ Myr (Sakamoto et al. 2006).   This indicate that the $l\!=\!1.3\deg$ region may be in an early stage of superbubble formation (`{\it proto-superbubble}').  We found a candidate of another proto-superbubble in the $l\!=\!-0.4\deg$ region, which should be investigated in more details.

\section{Discussion}
\subsection{Origin of High $R_{3\mbox{--}2/1\mbox{--}0}$ Gas}
In addition to the high \R32/10\ clumps, we also found a number of small spots of high \R32/10\ gas over the CMZ and Clump 2.  Many coincide with HVCCs, the other appear as high-velocity wings emanating from giant molecular clouds, suggesting that they are shocked molecular gas.  However, the spatial correlation with HVCCs is not excellent.  In other words, not all HVCCs contain large fraction of high \R32/10\ gas.  The total {\it J}=3--2/{\it J}=1--0 intensity ratio of HVCCs ranges from 0.4 to 2.2.  This variety of \R32/10\ could be due to evolution, or due to different origins of HVCCs.  Apparently, more research is necessary to solve this problem.  Except for the Sgr A and G 0.9+0.1 regions, high \R32/10\ gas in the Galactic center region seems to be predominantly shocked molecular gas.  Since the spatial distribution of high \R32/10\ gas is spotty and clumpy, its origin should be local explosive events, most likely supernova explosions.

\begin{figure}[!ht]


\begin{center}
\includegraphics[height=5cm]{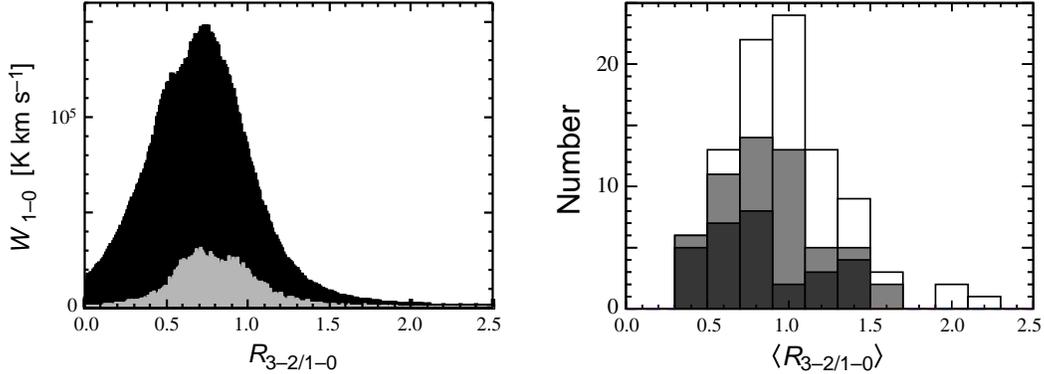}
\end{center}

\caption{({\it a}) Frequency distributions of \R32/10\ weighted by the CO {\it J}=1--0 intensity.  Grey area shows the HVCC data.  
({\it b}) Frequency distribution of the total {\it J}=3--2/{\it J}=1--0 intensity ratio of each HVCC.  White bars show HVCCs identified in the CO {\ it J}=3--2 data set only.  
}\label{fig5}
\end{figure}

\subsection{Stellar Clusters Associated with HVCCs}
Star formation activity in the recent past can be examined by the current supernova rate.  If each HVCC has been formed by a series of supernova explosions, it must be associated with a stellar cluster severely veiled in dense molecular material.  Age of such cluster must be younger than 30 million years, which is the main-sequence lifetime of an 8 $M_{\sun}$ star.  Assuming that each cluster was formed by a microburst of star formation, and assuming a Salpeter-type initial mass function ($dN/dM\!\propto\!M^{-1.65}$; Figer et al. 1999), we can estimate a cluster mass using the kinetic energy and expansion time of a HVCC.  The mass of currently dying stars was assumed to be 15 $M_{\sun}$.  The estimated cluster mass ranges from $10^2/\eta$ to $10^5/\eta\,M_{\sun}$, where $\eta$ is the conversion efficiency to the kinetic energy.  Note that some of them are very massive, exceeding $10^4/\eta\,M_{\sun}$.

\begin{figure}[!ht]
\begin{center}
\includegraphics[height=5cm]{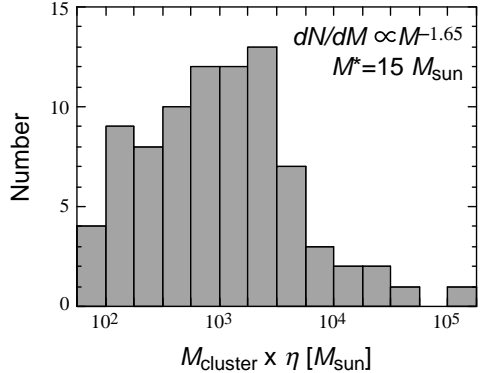}
\end{center}
\caption{Frequency distribution of cluster mass estimated from the kinetic energy and expansion time of HVCCs.  
}\label{fig6}
\end{figure}

\subsection{Turbulence Activation and Gas Heating}
Can supernova shocks contribute to turbulence activation and gas heating in the CMZ?  Here we consider the energy flow from large-scale to small-scale (Tanaka et al. 2009).  First, the energy injection rate by supernovae are estimated from kinetic energies and expansion times of HVCCs, to be $L_{\rm kin}\!=\!\sum E_{\rm kin}/t_{\rm exp}\!\sim\!10^{40}$ erg/s.  Second, the turbulence dissipation rate can be estimated from the typical velocity width ($\sigma_{\rm V}$) and size ($S$) of molecular clouds in the CMZ, to be $L_{\rm kin}\!=\!(1/2)M_{\rm CMZ}\,\sigma_{\rm V}^3/S\!\sim\!{\rm several}\!\times\!10^{39}$ erg/s.  And third, the gas cooling rate can be approximated by the cooling rate by CO rotational lines (Goldsmith \&\ Langer 1978).  Using a set of canonical values for the CMZ, $n(H_2)\!=\!10^4$ cm$^{-3}$, $T_{\rm k}\!=\!40$ K, $X_{\rm CO}/(dV/dR)\!=\!10^{-5\sim-4.5}$ pc (km s$^{-1}$)$^{-1}$, we had $L_{\rm CO}\!=\!(1\mbox{--}2)\!\times\!10^{39}$ erg/s.  These three values are comparable, and $L_{\rm kin}\!>\!L_{\rm D}\!>\!L_{\rm CO}$, suggesting that supernova shocks responsible for the HVCC formation can make significant contributions to turbulence activation and gas heating in the CMZ.

\section{Summary}
Our ASTE CO {\it J}=3--2 survey have revealed the distribution and kinematics of highly-excited gas in the Galactic center region.  We found a large infalling disk of high \R32/10\ gas in the Sgr A region, a candidate of proto-superbubble in the $l\!=\!+1.3\deg$ region, and a number of high \R32/10\ spots spreading over the CMZ and Clump2.  High \R32/10\ gas in the Galactic center region seems to be predominantly shocked molecular gas.  

The results also have revealed the nature of HVCCs.  The majority of HVCCs may have been formed by supernova explosions, and thus they must be associated with stellar clusters veiled in dense molecular material.  Cluster mass ranges from $10^2/\eta\,M_{\sun}$ to $10^5/\eta\,M_{\sun}$.  In the near future, relatives of HVCC will be found in centers of nearby galaxies with the Atacama Large Millimeter and submillimeter Array (ALMA).  

Rough estimates of energy injection rate, turbulence dissipation rate, and gas cooling rate gave comparable values, suggesting that supernova shocks responsible for the formation of HVCCs can provide significant contributions to turbulence activation and gas heating in the CMZ.

\acknowledgements 
We thank all members of the ASTE team for excellent performance of the telescope and every support in observations.   This study was financially supported by the MEXT Grant-in-Aid for Scientific Research on Priority Areas No.\ 15071202.  Observations with ASTE were in part carried out remotely from Japan by using NTT's GEMnet2 and its partner R\&E (Research and Education) networks, which are based on AccessNova collaboration of University of Chile, NTT Laboratories, and National Astronomical Observatory of Japan.


\end{document}